\documentclass[floats, eqnum, showpacs, nofootinbib, preprint,
eqsecnum]{revtex4-1}

\usepackage{color,graphicx}
\usepackage{amsfonts}
\usepackage{amssymb}
\begin{document}

\title{Retrolensing by a wormhole at deflection angles $\pi$ and $3\pi$}

\author{Naoki Tsukamoto}\email{tsukamoto@rikkyo.ac.jp}

\affiliation{School of Physics, Huazhong University of Science and Technology, Wuhan 430074, China}
\date{\today}

\begin{abstract}
The deflection angle of a light ray can be arbitrarily large near a light sphere. 
The time-symmetrical shape of light curves of a pair of light rays reflected by a light sphere of a lens object does not depend on the details of the lens object.
We consider retrolensing light curves of sunlight with deflection angles $\pi$ and $3\pi$ by an Ellis wormhole, which is the simplest Morris-Thorne wormhole.
If an Ellis wormhole with a throat parameter $a=10^{11}$ km is $100$ pc away from an observer 
and if the Ellis wormhole, the observer, and the sun are aligned perfectly in this order,
the apparent magnitudes of a pair of light rays with deflection angles $\pi$ and $3\pi$ become $11$ and $18$, respectively.
The two pairs of light rays make a superposed light curve with two separable peaks and they break down time symmetry of a retrolensing light curve.
The observation of the two separated peaks of the light curves gives us information on the details of the lens object.
If the observer can also separate the pair of the images with the deflection angle $\pi$ into a double image,
he or she can say whether the retrolensing is caused by an Ellis wormhole or a Schwarzschild black hole. 
\end{abstract}

\maketitle

\section{Introduction}
Gravitational lensing is a useful tool to search for dark and compact objects.
Gravitational lensing under a quasi-Newtonian approximation has been discussed widely~\cite{Schneider_Ehlers_Falco_1992,Petters_Levine_Wambsganss_2001,Schneider_Kochanek_Wambsganss_2006},
while gravitational lensing without the approximation has also been investigated~\cite{Perlick_2004_Living_Rev}.
In 1959 Darwin pointed out that an infinite number of ghost images appear near a light sphere or photon sphere~\cite{Claudel:2000yi,Hasse_Perlick_2002} in the Schwarzschild spacetime~\cite{Darwin_1959}.
The gravitational lensing of these faint images has been discussed by several authors~\cite{Atkinson_1965,Luminet_1979,Ohanian_1987,Nemiroff_1993,Frittelli_Kling_Newman_2000,Virbhadra_Ellis_2000,Bozza_Capozziello_Iovane_Scarpetta_2001,Bozza_2002,Perlick_2004_Phys_Rev_D,Iyer:2006cn,Bozza:2007gt,Bozza:2008ev,AzregAinou:2012xv,Ishihara:2016sfv,Tsukamoto:2016jzh}.

Gravitational lensing of light rays with a deflection angle which is almost $\pi$ is called retrolensing. 
Holz and Wheeler discussed the retrolensing of sunlight reflected by the light sphere of a black hole near the solar system~\cite{Holz:2002uf}.
One can distinguish between the light curves of retrolensing and other light curves since the retrolensing light curves have a characteristic shape and are symmetric in time.
If one observes a light curve with the characteristic shape and with solar spectra on the elliptic, 
he or she can say that it is a retrolensing light curve of the sun.
A black hole in the galactic center as a retrolens was also investigated~\cite{Eiroa:2003jf,Bozza:2004kq,DePaolis:2003ad}.
The effects of an electrical charge~\cite{Eiroa:2003jf,Tsukamoto:2016oca} and rotation~\cite{DePaolis:2003ad,DePaolis:2004xe} 
of a black hole on a retrolensing light curve and a double image~\cite{Tsukamoto:2016oca} were also considered.

Gravitational lensing is also caused by a wormhole~\cite{Kim_Cho_1994}.
One can survey wormholes with a negative Arnowitt-Deser-Misner (ADM) mass~\cite{Cramer:1994qj,Torres:1998xd,Anchordoqui:1999gca,Safonova:2001nd,Eiroa:2001zz,Safonova:2001vz,Safonova:2002si,Takahashi_Asada_2013},   
a vanishing ADM mass~\cite{Tsukamoto:2016jzh,Takahashi_Asada_2013,Chetouani_Clement_1984,Nandi_Zhang_Zakharov_2006,Dey_Sen_2008,Abe_2010,Bhattacharya:2010zzb,Toki_Kitamura_Asada_Abe_2011,Tsukamoto_Harada_2013,Kitamura_Nakajima_Asada_2013%
,Nakajima_Asada_2012,Tsukamoto_Harada_Yajima_2012,Gibbons_Vyska_2012,Yoo_Harada_Tsukamoto_2013,Lukmanova_2016,Tsukamoto:2016zdu,Tsukamoto:2016qro},
and a positive ADM mass~~\cite{Rahaman:2007am,Kuhfittig:2015sta,Nandi:2016ccg,Nandi_Zhang_Zakharov_2006,Tsukamoto:2016zdu,Tejeiro_Larranaga_2012,Sajadi:2016hko}
by gravitational lensing.
In 1973 Ellis obtained a static and spherically symmetric wormhole solution of Einstein equations with a phantom scalar field~\cite{Ellis_1973}. 
The traversable wormhole with vanishing ADM masses is called an Ellis wormhole or Ellis-Bronnikov wormhole 
since it was also obtained by Bronnikov in a scalar-tensor theory in the same year~\cite{Bronnikov_1973}.
The wormhole is often also referred to as a Morris-Thorne wormhole~\cite{Morris_Thorne_1988}
without mentioning Ellis's~\cite{Ellis_1973} and Bronnikov's works~\cite{Bronnikov_1973}.
The instability of the Ellis wormhole was revealed by several authors~\cite{Shinkai_Hayward_2002}, contrary to a conclusion of an earlier work~\cite{Armendariz-Picon_2002}.

Some static and spherically symmetric wormholes have the same metric as that of an Ellis wormhole 
in the vanishing ADM mass case~\cite{Kar:2002xa,Das:2005un,Shatskiy:2008us,Novikov:2012uj,Myrzakulov:2015kda}. 
In 2013 Bronnikov~\textit{et al.} showed that a wormhole with the same metric as the metric of an Ellis wormhole and with electrically charged dust with negative energy density 
is linearly stable under spherically symmetric and axial perturbations~\cite{Bronnikov:2013coa}. 
The quasinormal mode was also investigated~\cite{Konoplya:2016hmd}.

The trajectory of a light ray in the Ellis wormhole spacetime was investigated by Ellis in Ref.~\cite{Ellis_1973}. 
The deflection angle of the light ray was calculated first by Chetouani and Clement~\cite{Chetouani_Clement_1984} 
and then by several authors~\cite{Gibbons_Vyska_2012,Bhattacharya:2010zzb,Nakajima_Asada_2012,Tsukamoto_Harada_Yajima_2012,Tsukamoto:2016jzh,Tsukamoto:2016qro,Nandi_Zhang_Zakharov_2006,Dey_Sen_2008}.
Various gravitational lensing effects~\cite{Nandi_Zhang_Zakharov_2006,Tsukamoto_Harada_Yajima_2012,Tsukamoto:2016qro,Muller:2008zza%
,Yoo_Harada_Tsukamoto_2013,Lukmanova_2016,Abe_2010,Toki_Kitamura_Asada_Abe_2011,Tsukamoto_Harada_2013,Kitamura_Nakajima_Asada_2013,Tsukamoto:2016zdu%
,Izumi_Hagiwara_Nakajima_Kitamura_Asada_2013,Kitamura_Izumi_Nakajima_Hagiwara_Asada_2013,Tsukamoto:2014dta,Nakajima:2014nba,Bozza:2015wbw},
 a particle collision~\cite{Tsukamoto:2014swa}, a shadow~\cite{Ohgami:2015nra,Ohgami:2016iqm,Perlick:2015vta}, 
 visualization~\cite{Muller_2004}, quantum metrology~\cite{Sabin:2016zqo,Sabin:2017dvx},
and several observables like a rotation curve~\cite{Bozza:2015haa}  in the Ellis wormhole spacetime were also investigated.
Takahashi and Asada gave the upper bound of the number density $N_{upper} \leq 10^{-4}h^{3}\mathrm{Mpc}^{-3}$ 
of the Ellis wormhole with a throat parameter  $10 \leq a \leq 10^{4}$pc~\cite{Takahashi_Asada_2013}
with strong lensing of quasars in the data of the Sloan Digital Sky Survey Quasar Lens Search~\cite{Inada_Oguri_Shin_et_al_2012},
and Yoo \textit{et al.} gave $N_{upper} \leq 10^{-9}\mathrm{AU}^{-3}$ for $a\sim 1$cm~\cite{Yoo_Harada_Tsukamoto_2013}
with femtolensing of gamma-ray bursts~\cite{Barnacka_Glicenstein_Moderski_2012} 
in the data of the Fermi Gamma-Ray Burst Monitor~\cite{Meegan_Lichti_Bhat_et_al_2009}.

Recently, Tsukamoto and Harada made a conjecture that the shape of light curves formed by light rays which are reflected 
by a light sphere does not depend on the details of a static spherically symmetric and asymptotically flat spacetime and a lens configuration~\cite{Tsukamoto:2016zdu}.
If the conjecture is true, one cannot distinguish between black holes and wormholes with the shape of their retrolensing light curves.
Can we distinguish between black holes and wormholes by retrolensing?
In this paper, we consider the details of retrolensing by an Ellis wormhole and a black hole near the solar system to answer the question. 

This paper is organized as follows. 
In Sec.~II we briefly review a deflection angle in a strong deflection limit in an Ellis wormhole spacetime.
In Sec.~III we review retrolensing in a general static spherically symmetric and asymptotically flat spacetime. 
In Sec.~IV, we investigate the effect of the light rays with deflection angle $3\pi$ on retrolensing by the Ellis wormhole
and we discuss our results in Sec.~V.
In this paper we use the units in which the speed of light and Newton's constant are unity.

\section{Deflection angle in a strong deflection limit}
In this section, we review briefly the deflection angle $\alpha$ of a light ray in a strong deflection limit in an Ellis wormhole spacetime
and in the Schwarzschild spacetime in the following form,
\begin{equation}\label{eq:deflection_strong}
\alpha=-\bar{a} \log \left( \frac{b}{b_{c}}-1 \right) + \bar{b}+O((b-b_{c})\log(b-b_{c})),
\end{equation}
where $\bar{a}$ and $\bar{b}$ are parameters, $b$ is the impact parameter of the light ray, 
and $b_{c}$ is the critical impact parameter.~\footnote{%
Bozza estimated the order of the error term is $O(b-b_{c})$ in Ref.~\cite{Bozza_2002}
and then Tsukamoto pointed out that it should be read as $O((b-b_{c})\log(b-b_{c}))$~\cite{Tsukamoto:2016qro}.}

\subsection{Ellis wormhole}
The line element is given by~\cite{Ellis_1973,Bronnikov_1973}
\begin{equation}\label{eq:line_element}
ds^{2}=-dt^{2}+dr^{2}+(r^{2}+a^{2})(d\theta^{2}+\sin^{2}\theta d\phi^{2}),
\end{equation}
where $a$ is a positive constant.
The wormhole throat exists at $r=0$.
The coordinates are defined in a range $-\infty < t < \infty, -\infty < r < \infty, 0 \leq \theta  \leq \pi$, and $0 \leq \phi < 2\pi$
but we concentrate on a region $r\geq 0$.
We assume $\theta=\pi/2$ without loss of generality because of spherical symmetry.

From $ds^{2}=0$,
the trajectory of a light ray is given by
\begin{equation}\label{eq:trajectory}
\frac{1}{(r^{2}+a^{2})^{2}} \left( \frac{dr}{d\phi} \right)^{2}= \frac{1}{b^{2}}-\frac{1}{r^{2}+a^{2}},
\end{equation}
where $b\equiv L/E$ is the impact parameter of the light ray.
The conserved energy $E\equiv \dot{t}>0$ and angular momentum $L\equiv (r^{2}+a^{2})\dot{\phi}$, 
where the dot denotes differentiation with respect to an affine parameter, are constant along the trajectory.
A light ray does not pass through the throat if $\left| b \right| >a$
while it passes through the throat if $\left| b \right| <a$.
We only consider $\left| b \right| >a$.
The Ellis wormhole spacetime has a light sphere at $r=0$ which is coincident with the throat.
In the strong deflection limit $b\rightarrow b_{c}\equiv a$, where $b_{c}$ is the critical impact parameter,
light rays wind around the wormhole throat at $r=0$.
 
From Eq.~(\ref{eq:trajectory}), the deflection angle of a light ray is given by~\cite{Chetouani_Clement_1984}
\begin{eqnarray}\label{eq:deflection_angle1}
\alpha(b)
&=&2\int^{\infty}_{r_{o}} \frac{b dr}{\sqrt{(r^{2}+a^{2})(r^{2}+a^{2}-b^{2})}}-\pi \nonumber\\
&=&2K \left( \frac{a}{b} \right) -\pi,
\end{eqnarray}
where $r_{o}\equiv \sqrt{b^{2}-a^{2}}$ is the closest distant of the light ray
and $K(k)$ is the complete elliptic integral of the first kind given by
\begin{equation}
K(k)=\int^{1}_{0} \frac{dx}{\sqrt{(1-x^{2})(1-k^{2}x^{2})}},
\end{equation}
where $0<k<1$.
In the strong deflection limit $b\rightarrow b_{c}=a$, 
the deflection angle becomes
\begin{equation}
\alpha (b)=-\log \left( \frac{b}{b_{c}}-1 \right) +3\log 2-\pi +O((b-b_{c})\log(b-b_{c})).
\end{equation}
Thus, $\bar{a}=1$ and $\bar{b}=3\log 2-\pi$ in Eq.~(\ref{eq:deflection_strong}).
Here we have used 
\begin{equation}
\lim_{k\rightarrow 1}K(k)=-\frac{1}{2}\log(1-k)+\frac{3}{2}\log2 +O((1-k)\log(1-k)),
\end{equation}
which is obtained from Eq.~(10) in section 13.~8 in Ref.~\cite{Erdelyi}.
See Ref.~\cite{Tsukamoto:2016qro} for the details of the deflection angle in the strong deflection limit in an Ellis wormhole spacetime.

\subsection{Schwarzschild black hole}
The line element in the Schwarzschild spacetime is given by
\begin{equation}
ds^{2}=-\left(1-\frac{2M}{r}\right)dt^{2}+\frac{dr^{2}}{1-\frac{2M}{r}}+r^{2}(d\theta^{2}+\sin^{2}\theta d\phi^{2}),
\end{equation}
where $M$ is the ADM mass.
The critical impact parameter of a light ray is given by $b_{c}\equiv 3\sqrt{3}M$~\cite{Darwin_1959,Bozza_2002,Iyer:2006cn}.
The deflection angle $\alpha$ of the light ray in the strong deflection limit $b \rightarrow b_{c}$ is expressed as~\cite{Darwin_1959,Bozza_2002,Iyer:2006cn,Tsukamoto:2016jzh}
\begin{eqnarray}
\alpha (b)
&=&-\log \left( \frac{b}{b_{c}}-1 \right) +\log \left[ 216(7-4\sqrt{3}) \right] -\pi \nonumber\\
&&+O((b-b_{c})\log(b-b_{c})).
\end{eqnarray}
Thus, we obtain $\bar{a}=1$ and $\bar{b}=\log \left[ 216(7-4\sqrt{3}) \right]-\pi$.

\section{Retrolensing}
In this section, we review retrolensing~\cite{Holz:2002uf,Eiroa:2003jf,Bozza:2004kq,Tsukamoto:2016oca} 
in a general static spherically symmetric and asymptotically flat spacetime.

\subsection{Lens equation}
\begin{figure}[htbp]
\begin{center}
\includegraphics[width=70mm]{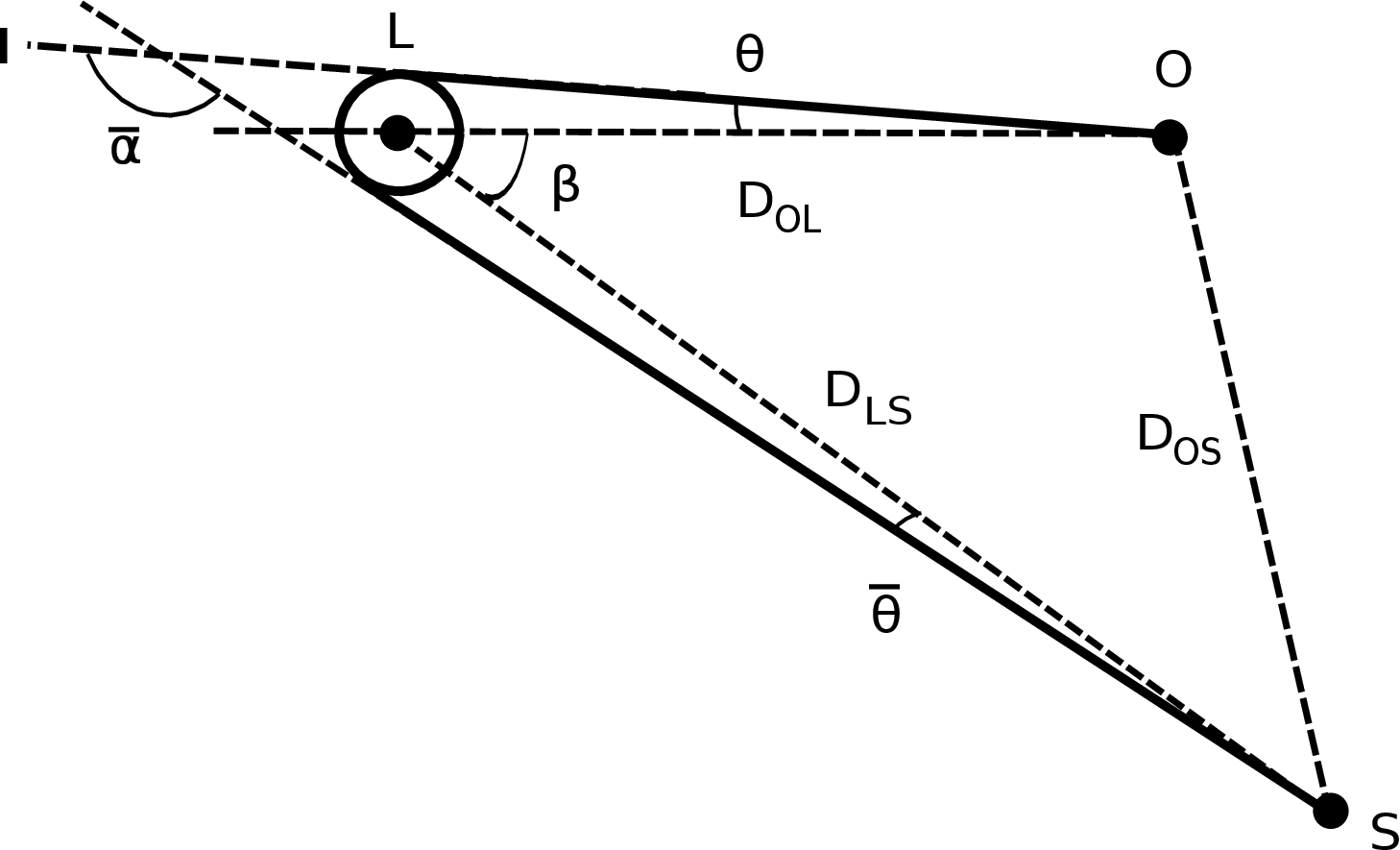}
\end{center}
\caption{Retrolensing configuration.
A light ray emitted by the sun $S$ is deflected by a lens $L$ or a light sphere with an deflection angle $\alpha$.
An observer $O$ sees an image $I$ with an image angle $\theta$. 
$\bar{\alpha}$ is the effective deflection angle of the light ray which rotates $N$ times around the light sphere 
defined as $\bar{\alpha}\equiv \alpha-2\pi N$.
$\bar{\theta}$ is an angle at $S$ between the light ray and line $LS$ and 
$\beta$ is a source angle defined by $\angle OLS$.
$D_{OL}$, $D_{LS}$, and $D_{OS}$ are the distances between the observer and the lens, 
between the lens and the source, and between the observer and the source, respectively.
}
\label{Lens_Configuration}
\end{figure}

We consider that a light ray emitted by the sun $S$ is deflected by a lens $L$ or a light sphere with an deflection angle $\alpha$ 
and then it reaches an observer $O$ with an image angle $\theta$. 
Figure~\ref{Lens_Configuration} shows the retrolensing configuration.
We concentrate on the case where the impact parameter $b$ is positive.
We define the effective deflection angle $\bar{\alpha}$ of the light ray as 
\begin{equation}\label{eq:effective_deflection_angle}
\bar{\alpha}\equiv \alpha-2\pi N
\end{equation}
where $N$ is a nonnegative integer which denotes the winding number of the light ray around the light sphere.
We use a lens equation considered by Ohanian~\cite{Ohanian_1987,Bozza:2008ev},
\begin{equation}\label{eq:lens}
\beta=\pi-\bar{\alpha}(\theta)+\theta+\bar{\theta},
\end{equation}
where $\beta$ is a source angle $\angle OLS$ and $\bar{\theta}$ is an angle between the light ray and the line $LS$ at $S$.
We assume that the lens $L$, the observer $O$, and the sun $S$ are almost aligned in this order 
and that the size of the light sphere is apparently small for the observer, i.e., 
\begin{equation}
b_{c} \ll D_{OL},
\end{equation}
where $D_{OL}$ is the distance between the lens and the observer.
Under these assumptions, we obtain
\begin{equation}
\alpha\sim \pi+2\pi N,
\end{equation}
\begin{equation}
\bar{\alpha}\sim \pi,
\end{equation}
\begin{equation}
\beta \sim 0,
\end{equation}
\begin{equation}
D_{LS}=D_{OL}+D_{OS},
\end{equation}
where $D_{LS}$ and $D_{OS}$ are the distances between the lens and source and between the observer and the source, respectively, 
\begin{equation}\label{eq:b_theta}
\frac{b}{D_{OL}}=\theta \sim \frac{b_{c}}{D_{OL}} \ll 1,
\end{equation}
and
\begin{equation}
\bar{\theta} \sim \frac{b_{c}}{D_{LS}} \ll 1. 
\end{equation}
Under these assumptions, we can justify neglecting the terms $\theta$ and $\bar{\theta}$ in the Ohanian lens equation~(\ref{eq:lens}).

\subsection{Image angle}
Inserting the deflection angle in the strong deflection limit~(\ref{eq:deflection_strong}), 
the definition of the effective deflection angle~(\ref{eq:effective_deflection_angle}), 
and Eq.~(\ref{eq:b_theta})
into the Ohanian lens equation~(\ref{eq:lens}) and neglecting the terms $\theta$ and $\bar{\theta}$,
we obtain positive solutions $\theta=\theta_{+N}(\beta)$ for every $N$, as obtained in Ref.~\cite{Bozza:2004kq},
where
\begin{equation}\label{eq:theta+N}
\theta_{+N}(\beta) \equiv \frac{b_{c}}{D_{OL}} \left( 1+e_{+N}(\beta) \right),
\end{equation}
where $e_{+N}(\beta)$ is defined as
\begin{eqnarray}
e_{+N}(\beta) 
&\equiv& \exp \left[ \frac{\bar{b}-(1+2N)\pi +\beta}{\bar{a}} \right] \nonumber\\
&\sim& \exp \left[ \frac{\bar{b}-(1+2N)\pi}{\bar{a}} \right].
\end{eqnarray}
When $N=0$, we obtain
\begin{eqnarray}
\theta_{+0}(\beta)
&=& \frac{b_{c}}{D_{OL}} \left[ 1+\exp \left( \frac{\bar{b}-\pi +\beta}{\bar{a}} \right) \right] \nonumber\\
&\sim& \frac{b_{c}}{D_{OL}} \left[ 1+\exp \left( \frac{\bar{b}-\pi }{\bar{a}} \right) \right].
\end{eqnarray}

From spherical symmetry, the negative solution $\theta=\theta_{-N}(\beta)$ of the Ohanian lens equation 
denoting the image angle of a light ray with a negative impact parameter is given by
\begin{equation}
\theta_{-N}(\beta) =-\theta_{+N}(-\beta)\sim -\theta_{+N}(\beta)
\end{equation}
for each $N$.
The image separation $\Delta \theta_{N}$ between the positive and negative images for every nonnegative integer $N$ is obtained as
\begin{eqnarray}
\Delta \theta_{N}
\equiv \theta_{+N}(\beta)-\theta_{-N}(\beta)
\sim 2\theta_{+N}(\beta).
\end{eqnarray}

\subsection{Magnification}
We assume that the sun is regarded as a uniform-luminous disk~\cite{Witt:1994,Nemiroff:1994uz,Alcock:1997fi} with a radius $R_{s}$, where $R_{s}=7\times10^{5}$km is the radius of the sun.
The magnification $\mu_{+N}(\beta)$ of the image with $\theta_{+N}(\beta)$ is obtained as~\cite{Bozza:2004kq,Tsukamoto:2016oca}
\begin{equation}\label{eq:mu+N}
\mu_{+N}(\beta)=-\frac{2D_{OS}^{2}\theta_{+N}}{\pi D_{LS}R_{s}^{2}} \frac{d\theta_{+N}}{d\beta} I(\beta),
\end{equation}
where $I(\beta)$ is given by, for $\beta D_{LS} \leq R_{s}$,
\begin{eqnarray}
I(\beta)
&=&\pi(R_{s}-\beta D_{LS}) \nonumber\\
&&+\int^{R_{s}+\beta D_{LS}}_{R_{s}-\beta D_{LS}} 
\arccos \frac{-R_{s}^{2}+\beta^{2} D_{LS}^{2}+R^{2}}{2\beta D_{LS}R}dR \nonumber\\
\end{eqnarray}
and, for $R_{s}\leq \beta D_{LS}$,
\begin{equation}
I(\beta)
=\int^{R_{s}+\beta D_{LS}}_{-R_{s}+\beta D_{LS}} 
\arccos \frac{-R_{s}^{2}+\beta^{2} D_{LS}^{2}+R^{2}}{2\beta D_{LS}R}dR. 
\end{equation}
From Eqs.~(\ref{eq:theta+N}) and (\ref{eq:mu+N}), the magnification is obtained as 
\begin{equation}\label{eq:mu+N2}
\mu_{+N}(\beta)=-\frac{2D_{OS}^{2}b_{c}^{2}}{\pi D_{LS}D_{OL}^{2}R_{s}^{2}} \frac{e_{+N}(\beta)\left( 1+e_{+N}(\beta) \right)}{\bar{a}} I(\beta).
\end{equation}

The magnification $\mu_{-N}(\beta)$ of the image with image angle $\theta_{-N}(\beta)$ is given by 
\begin{equation}\label{eq:mu-N}
\mu_{-N}(\beta)\sim -\mu_{+N}(\beta).
\end{equation}
The total magnification $\mu_{totN}(\beta)$ of a pair of images for each $N$ is given as
\begin{eqnarray}\label{eq:mutotN}
\mu_{totN}(\beta)
&\equiv& \left|\mu_{+N}(\beta)\right| +\left|\mu_{-N}(\beta)\right| \nonumber\\
&=&\frac{4D_{OS}^{2}b_{c}^{2}}{\pi D_{LS}D_{OL}^{2}R_{s}^{2}} \frac{e_{+N}(\beta)\left( 1+e_{+N}(\beta) \right)}{\bar{a}} \left| I(\beta) \right|. \nonumber\\
\end{eqnarray}
In a perfectly aligned case, $\beta=0$, the total magnification becomes
\begin{equation}
\mu_{totN}(0)
=\frac{4D_{OS}^{2}b_{c}^{2}}{D_{LS}D_{OL}^{2}R_{s}} \frac{e_{+N}(0)\left( 1+e_{+N}(0) \right)}{\bar{a}}
\end{equation}
for every $N$. Here we have used $I(0)=\pi R_{s}$.

\subsection{Source plane}
\begin{figure}[htbp]
\begin{center}
\includegraphics[width=60mm]{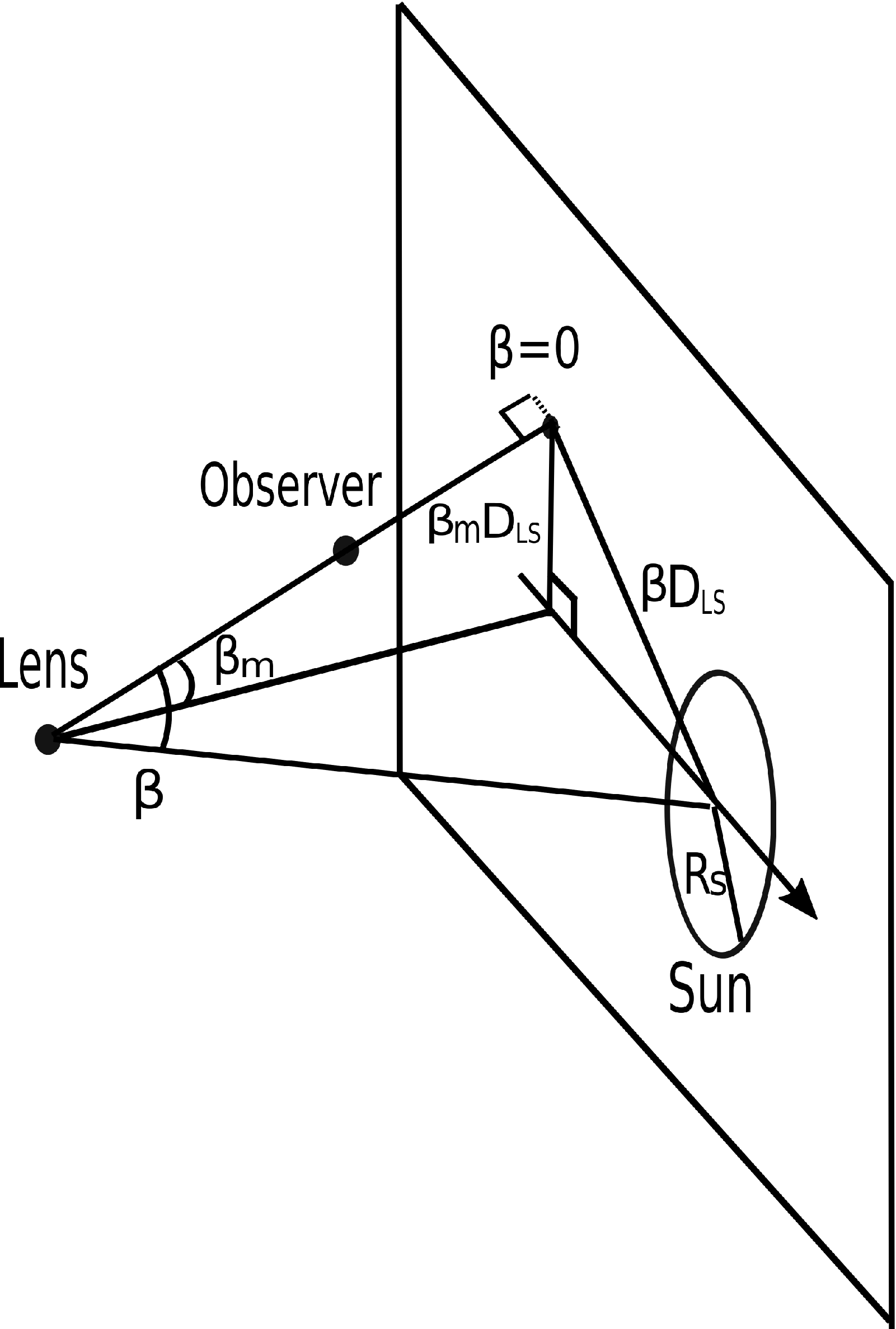}
\end{center}
\caption{Retrolensing. A source plane is defined as the plane that the sun is on and that is orthogonal to the optical axis, an axis $\beta=0$.
The closest separation $\beta_{m}$ is defined by the smallest source angle during retrolensing. 
We assume that the sun moves with the orbital velocity of the sun $v=30$km/s on the source plane.  
}
\label{Retrolensing_motion}
\end{figure}

The sun is on a source plane defined as a plane that is orthogonal to the optical axis, i.e., an axis $\beta=0$. 
We denote the closest separation between the center of the sun and the intersection of the source plane and the optical axis by $\beta_{m}$.
Figure~\ref{Retrolensing_motion} shows the source plane.

\subsection{Light curves}
\begin{figure}[htbp]
\begin{center}
\includegraphics[width=70mm]{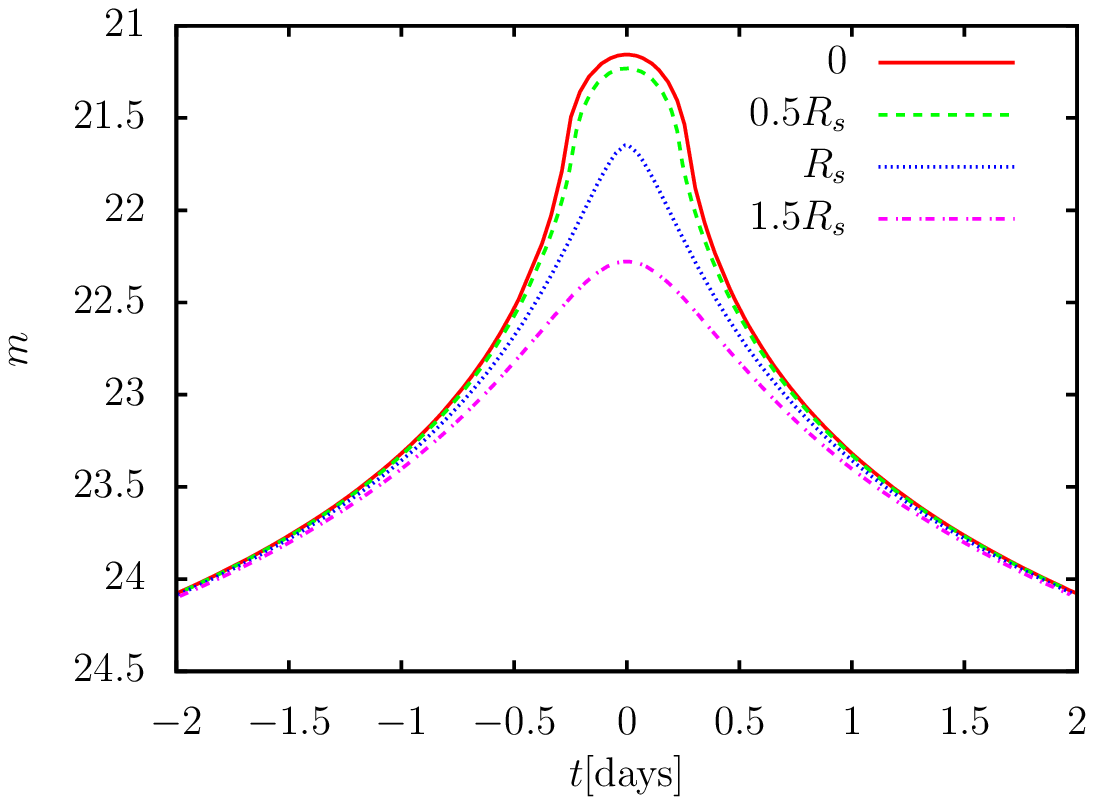}
\end{center}
\caption{Retrolensing light curves by an Ellis wormhole with $a=10^{6}$km at $D_{OL}=1$pc.
The (red) solid, (green) dashed, (blue) dotted, and (purple) dashed-dotted curves denote the light curves with 
$\beta_{m}D_{LS}=0$, $0.5R_{s}$, $R_{s}$, and $1.5R_{s}$, respectively.
$m$ is the apparent magnitude.}
\label{beta}
\end{figure}

The retrolensing light curves do not diverge because of the finite size of the sun.
Figure~\ref{beta} shows the retrolensing light curves by an Ellis wormhole with
$a=10^{6}$km at $D_{OL}=1$pc away.
One can estimate $\beta_{m}D_{LS}/R_{s}$ from the shape of the peak of the light curves 
since the light curves have a characteristic shape depending on $\beta_{m}D_{LS}/R_{s}$.
The characteristic time scale of the peak is obtained as $2R_{s}/v=12$~hours.
In general, if a lensing object is static and has spatial spherical symmetry, 
the whole shape and time scale of the retrolensing light curves of the sun do not depend on the parameters of the lens 
such as the ADM mass, the electrical charge, the size of the wormhole throat, and $D_{LS}$~\cite{Tsukamoto:2016oca,Tsukamoto:2016zdu}
while the peak shape strongly depends on $\beta_{m}D_{LS}/R_{s}$.

\begin{figure}[htbp]
\begin{center}
\includegraphics[width=70mm]{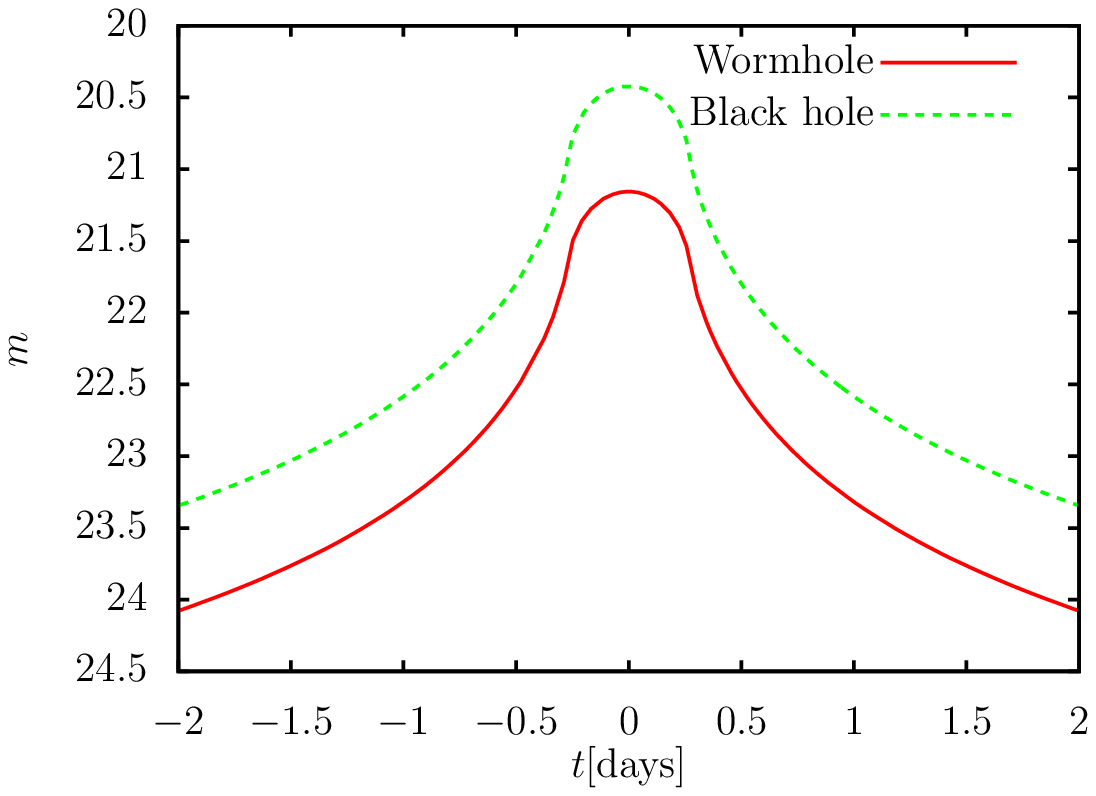}
\end{center}
\caption{Retrolensing light curves by an Ellis wormhole and a Schwarzschild black hole with $b_{c}=10^{6}$km and $\beta_{m}=0$ at $D_{OL}=1$pc.
The (red) solid and (green) dashed curves denote retrolensing light curves by
the wormhole and the black hole, respectively.
}
\label{whbh}
\end{figure}

Figure~\ref{whbh} shows the retrolensing light curves by an Ellis wormhole and a Schwarzschild black hole with $b_{c}=10^{6}$km and $\beta_{m}=0$ at $D_{OL}=1$pc.
We notice that the shapes of the retrolensing light curves are similar while the apparent magnitudes $m$ are different.

\section{Higher order effect: $N=1$}
In this section, we investigate the effect of the light rays with the deflection angle $\alpha \sim 3\pi$, 
i.e., the light rays with the winding number $N=1$,
on retrolensing light curves. 
The light rays with $N=1$ make a separable second peak of the retrolensing light curves if the light sphere of a lens object is large enough.

\subsection{Magnification with $N=1$}
The total magnification $\mu_{tot1}$ of the images with $N=1$ is obtained from Eq.~(\ref{eq:mutotN}).
The ratio between the total magnifications with $N=0$ and with $N=1$ is given by
\begin{equation}
\frac{\mu_{tot1}}{\mu_{tot0}} 
= \frac{   1+\exp \left( \frac{\bar{b}-3\pi}{\bar{a}} \right)  }
{ \exp \left( \frac{2\pi}{\bar{a}} \right)   \left[    1+\exp \left( \frac{\bar{b}-\pi}{\bar{a}} \right) \right] }
\sim \exp \left( -\frac{2\pi}{\bar{a}} \right).
\end{equation}
Notice that the ratio of the total magnifications only depends on the metric tensor.
The ratios of the total magnifications of retrolensing by the Ellis wormhole and the Schwarzschild black hole are given by 
$\mu_{tot1}/\mu_{tot0}=1.840 \times 10^{-3}$ and $1.815 \times 10^{-3}$, respectively.   
The difference of the ratio of the total magnifications $\mu_{tot1}/\mu_{tot0}$ by the Ellis wormhole and the Schwarzschild black hole is very small
since the parameter $\bar{a}$ is unity in both cases.

\subsection{Time delay between the light rays with $N=0$ and $N=1$}
Light rays with $N=1$ reach the observer after light rays with $N=0$ reach him or her.
The time delay $\Delta T$ of the light rays with $N=1$ compared to the light rays with $N=0$ is given by the proper length around the light sphere.
Thus, the time delay $\Delta T$ only depends on the metric tensor 
and its parameters such as the ADM mass $M$ and the throat parameter $a$.
The time delays $\Delta T$ in the Ellis wormhole spacetime and the Schwarzschild black hole spacetime are 
obtained as $\Delta T=2\pi a$ and $\Delta T=6\pi M$, respectively.

\subsection{Image separation of images with $N=0$ and $N=1$}
Image separation $\Delta \theta_{01}$ between images with $N=0$ and with $N=1$ is given by, from Eq.~(\ref{eq:theta+N}),
\begin{eqnarray}
\Delta \theta_{01}(\beta) 
&\equiv& \theta_{+0}-\theta_{+1}
\sim \theta_{+0}-\theta_{+\infty} \nonumber\\
&\sim& \frac{b_{c}}{D_{OL}} \exp \left( \frac{\bar{b}-\pi }{\bar{a}} \right),
\end{eqnarray}
where $\theta_{+\infty}$ is the image angle of the innermost image among an infinite number of images 
and it corresponds with the apparent size of the light sphere, 
\begin{equation}
\theta_{+\infty}= \frac{b_{c}}{D_{OL}}.
\end{equation}

\subsection{Light curves of two pairs of light rays with $N=0$ and $N=1$}
\begin{figure}[htbp]
\begin{center}
\includegraphics[width=70mm]{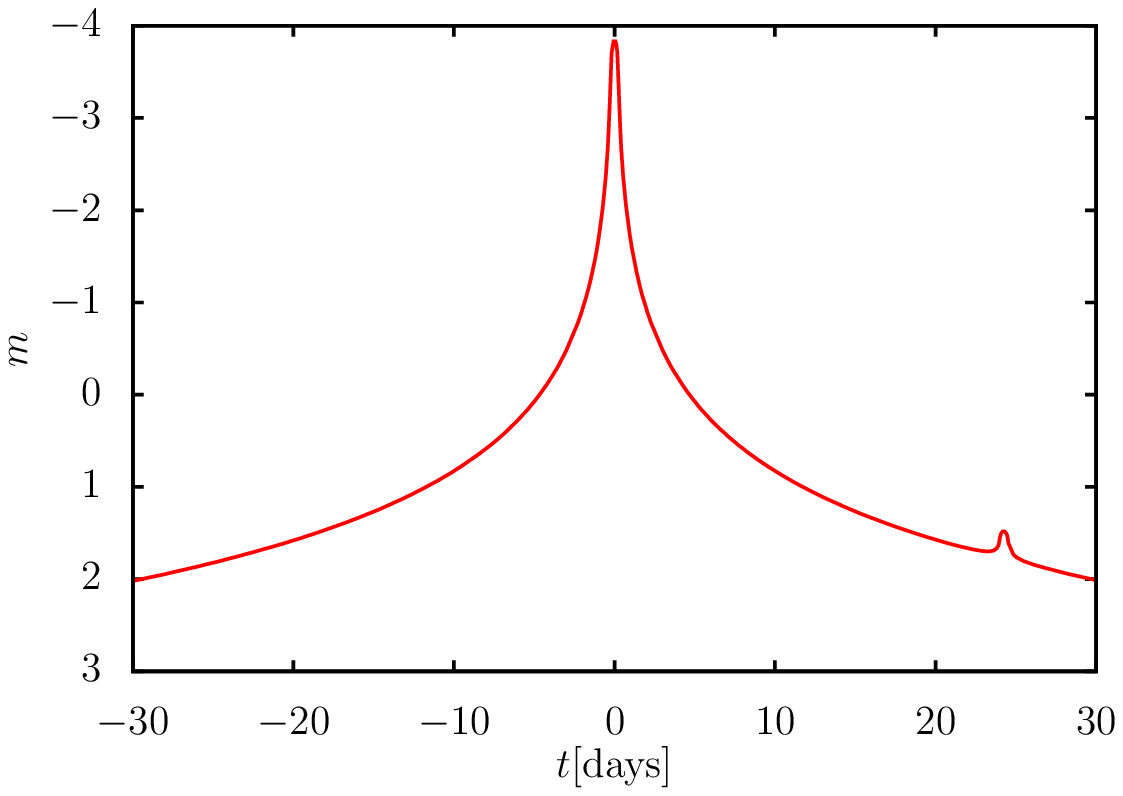}
\end{center}
\caption{Retrolensing light curves by an Ellis wormhole with $a=10^{11}$km and $\beta_{m}=0$ at $D_{OL}=1$pc.
}
\label{14m1pc}
\end{figure}

\begin{figure}[htbp]
\begin{center}
\includegraphics[width=70mm]{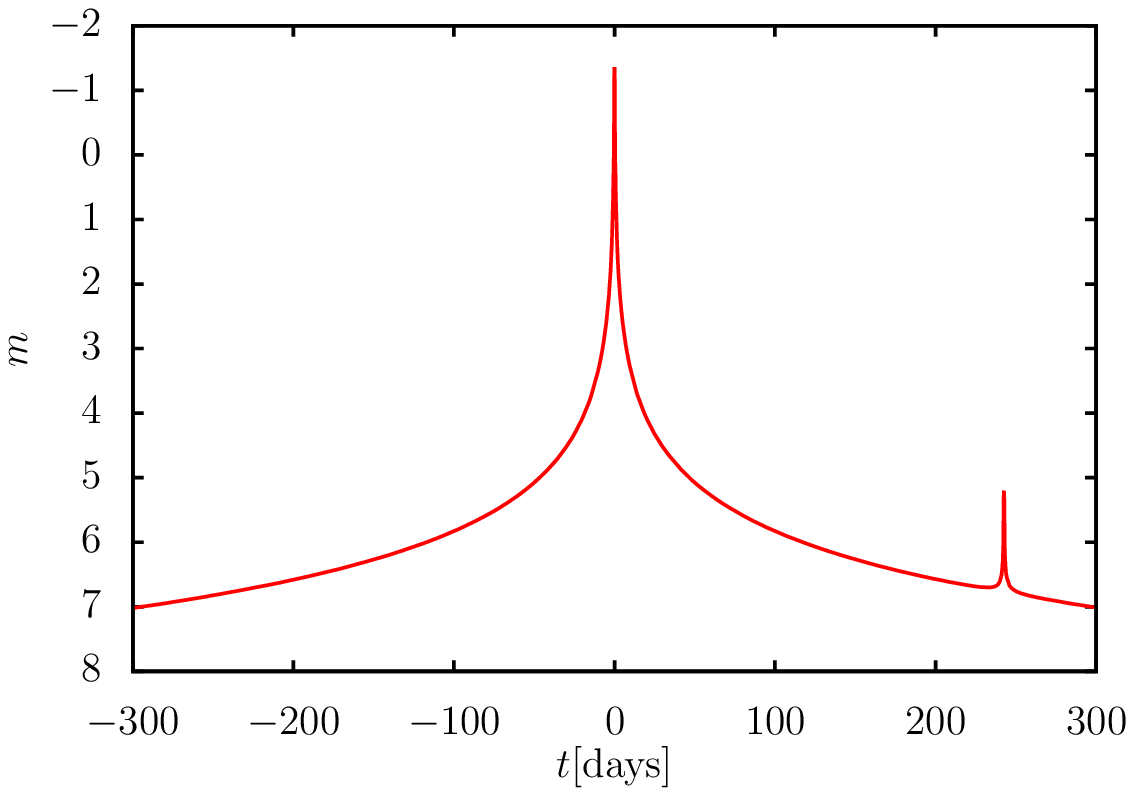}
\end{center}
\caption{Retrolensing light curves by an Ellis wormhole with $a=10^{12}$km and $\beta_{m}=0$ at $D_{OL}=10$pc.
}
\label{15m10pc}
\end{figure}

Figure~\ref{14m1pc} (Figure~\ref{15m10pc}) shows the light curves of two pairs of light rays with $N=0$ and $N=1$, 
which are retrolensed by an Ellis wormhole with $a=10^{11}$km at $D_{OL}=1$pc ($a=10^{12}$km at $D_{OL}=10$pc) in the perfectly aligned case, $\beta_{m}=0$.
A second small peak appears and time symmetry of the retrolensing light curves is broken because of the light rays with $N=1$.

Tables~\ref{table:I} shows the critical impact parameter $b_{c}$; 
the apparent magnitude $m_{0}$ of the sum of the pair of light rays with $N=0$ in the perfectly aligned case, i.e., $\beta=0$;
the apparent magnitude $m_{1}$ of the sum of the pair of light rays with $N=1$ in the perfectly aligned case;
the time delay $\Delta T$ of the light rays with $N=1$ comparing the light rays with $N=0$;
the image separation $\Delta \theta_{0}$ of the pair of light rays with $N=0$;
and the image separation $\Delta \theta_{01}$ between the images with $N=0$ and $N=1$ 
of retrolensing by an Ellis wormhole at $D_{OL}=1$pc.
Tables~\ref{table:II} and III~\ref{table:III} show the cases of retrolensing by an Ellis wormhole at $D_{OL}=10$pc 
and a Schwarzschild black hole at $D_{OL}=1$pc, respectively.
\begin{table}[hbtp]
 \caption{Retrolensing by an Ellis wormhole at $D_{OL}=1$pc.
$b_{c}$ is the critical impact parameter, 
$m_{0}$ ($m_{1}$) is the apparent magnitude of the total of the pair of light rays with $N=0$ ($N=1$) in the perfectly aligned case, 
$\Delta T$ is the time delay of the light rays with $N=1$ comparing the light rays with $N=0$,
$\Delta \theta_{0}$ is the image separation of the pair of the light rays with $N=0$, and
$\Delta \theta_{01}$ is the image separation between the images with $N=0$ and $N=1$.
 }
 \label{table:I}
\begin{center}
\begin{tabular}{c c c c c c} \hline
$b_{c}$(km)$\;$ &$m_{0}\;$ &$m_{1}\;$ &$\Delta T$(s)    $\;$ &$\Delta \theta_{0}$(mas) &$\Delta \theta_{01}$(mas) \\ \hline
$1          \;$ &$51     \;$ &$58     \;$ &$2.1\times 10^{-5}\;$ &$1.4\times 10^{-5}$ &$1.0\times 10^{-7}$            \\ \hline
$10         \;$ &$46     \;$ &$53     \;$ &$2.1\times 10^{-4}\;$ &$1.4\times 10^{-4}$ &$1.0\times 10^{-6}$           \\ \hline
$10^{2}     \;$ &$41     \;$ &$48     \;$ &$2.1\times 10^{-3}\;$ &$1.4\times 10^{-3}$ &$1.0\times 10^{-5}$ 	     \\ \hline
$10^{3}     \;$ &$36     \;$ &$43     \;$ &$2.1\times 10^{-2}\;$ &$1.4\times 10^{-2}$ &$1.0\times 10^{-4}$	     \\ \hline
$10^{4}     \;$ &$31     \;$ &$38     \;$ &$2.1\times 10^{-1}\;$ &$1.4\times 10^{-1}$ &$1.0\times 10^{-3}$ 	     \\ \hline
$10^{5}     \;$ &$26     \;$ &$33     \;$ &$2.1              \;$ &$1.4              $ &$1.0\times 10^{-2}$	     \\ \hline
$10^{6}     \;$ &$21     \;$ &$28     \;$ &$2.1\times 10     \;$ &$1.4\times 10     $ &$1.0\times 10^{-1}$ 	     \\ \hline
$10^{7}     \;$ &$16     \;$ &$23     \;$ &$2.1\times 10^{2} \;$ &$1.4\times 10^{2}$  &$1.0$ 	     \\ \hline
$10^{8}     \;$ &$11     \;$ &$18     \;$ &$2.1\times 10^{3} \;$ &$1.4\times 10^{3}$  &$1.0\times 10$ 	     \\ \hline
$10^{9}     \;$ &$6.2    \;$ &$13     \;$ &$2.1\times 10^{4} \;$ &$1.4\times 10^{4}$  &$1.0\times 10^{2}$ 	     \\ \hline
$10^{10}    \;$ &$1.2    \;$ &$8.0    \;$ &$2.1\times 10^{5} \;$ &$1.4\times 10^{5}$  &$1.0\times 10^{3}$ 	     \\ \hline
$10^{11}    \;$ &$-3.8   \;$ &$3.0    \;$ &$2.1\times 10^{6} \;$ &$1.4\times 10^{6}$  &$1.0\times 10^{4}$ 	     \\ \hline
\end{tabular}
\end{center}
\end{table}
\begin{table}[hbtp]
 \caption{Retrolensing by an Ellis wormhole at $D_{OL}=10$pc.}
 \label{table:II}
\begin{center}
\begin{tabular}{c c c c c c} \hline
$b_{c}$[km]$\;$ &$m_{0}\;$ &$m_{1}\;$ &$\Delta T$[s]    $\;$ &$\Delta \theta_{0}$[mas] &$\Delta \theta_{01}$[mas] \\ \hline
$1          \;$ &$59     \;$ &$65     \;$ &$2.1\times 10^{-5}\;$ &$1.4\times 10^{-6}$ &$1.0\times 10^{-8}$           \\ \hline
$10         \;$ &$54     \;$ &$60     \;$ &$2.1\times 10^{-4}\;$ &$1.4\times 10^{-5}$ &$1.0\times 10^{-7}$           \\ \hline
$10^{2}     \;$ &$49     \;$ &$55     \;$ &$2.1\times 10^{-3}\;$ &$1.4\times 10^{-4}$ &$1.0\times 10^{-6}$           \\ \hline
$10^{3}     \;$ &$44     \;$ &$50     \;$ &$2.1\times 10^{-2}\;$ &$1.4\times 10^{-3}$ &$1.0\times 10^{-5}$	     \\ \hline
$10^{4}     \;$ &$39     \;$ &$45     \;$ &$2.1\times 10^{-1}\;$ &$1.4\times 10^{-2}$ &$1.0\times 10^{-4}$ 	     \\ \hline
$10^{5}     \;$ &$34     \;$ &$40     \;$ &$2.1              \;$ &$1.4\times 10^{-1}$ &$1.0\times 10^{-3}$	     \\ \hline
$10^{6}     \;$ &$29     \;$ &$35     \;$ &$2.1\times 10     \;$ &$1.4              $ &$1.0\times 10^{-2}$ 	     \\ \hline
$10^{7}     \;$ &$24     \;$ &$30     \;$ &$2.1\times 10^{2} \;$ &$1.4\times 10$   	  &$1.0\times 10^{-1}$           \\ \hline
$10^{8}     \;$ &$19     \;$ &$25     \;$ &$2.1\times 10^{3} \;$ &$1.4\times 10^{2}$  &$1.0$ 	     \\ \hline
$10^{9}     \;$ &$14     \;$ &$20     \;$ &$2.1\times 10^{4} \;$ &$1.4\times 10^{3}$  &$1.0\times 10$ 	     \\ \hline
$10^{10}    \;$ &$8.7    \;$ &$15     \;$ &$2.1\times 10^{5} \;$ &$1.4\times 10^{4}$  &$1.0\times 10^{2}$ 	     \\ \hline
$10^{11}    \;$ &$3.7    \;$ &$10     \;$ &$2.1\times 10^{6} \;$ &$1.4\times 10^{5}$  &$1.0\times 10^{3}$ 	     \\ \hline
$10^{12}    \;$ &$-1.3   \;$ &$5.5    \;$ &$2.1\times 10^{7} \;$ &$1.4\times 10^{6}$  &$1.0\times 10^{4}$ 	     \\ \hline
\end{tabular}
\end{center}
\end{table}
\begin{table}[hbtp]
 \caption{Retrolensing by a black hole at $D_{OL}=1$pc.}
 \label{table:III}
\begin{center}
\begin{tabular}{c c c c c c} \hline
$b_{c}$[km]$\;$ &$m_{0}\;$ &$m_{1}\;$ &$\Delta T$[s]    $\;$ &$\Delta \theta_{0}$[mas] &$\Delta \theta_{01}$[mas] \\ \hline
$1          \;$ &$50     \;$ &$57     \;$ &$1.2\times 10^{-5}\;$ &$1.4\times 10^{-5}$ &$1.9\times 10^{-7}$       \\ \hline
$10         \;$ &$45     \;$ &$52     \;$ &$1.2\times 10^{-4}\;$ &$1.4\times 10^{-4}$ &$1.9\times 10^{-6}$           \\ \hline
$10^{2}     \;$ &$40     \;$ &$47     \;$ &$1.2\times 10^{-3}\;$ &$1.4\times 10^{-3}$ &$1.9\times 10^{-5}$ 	     \\ \hline
$10^{3}     \;$ &$35     \;$ &$42     \;$ &$1.2\times 10^{-2}\;$ &$1.4\times 10^{-2}$ &$1.9\times 10^{-4}$	     \\ \hline
$10^{4}     \;$ &$30     \;$ &$37     \;$ &$1.2\times 10^{-1}\;$ &$1.4\times 10^{-1}$ &$1.9\times 10^{-3}$           \\ \hline
$10^{5}     \;$ &$25     \;$ &$32     \;$ &$1.2              \;$ &$1.4              $ &$1.9\times 10^{-2}$	     \\ \hline
$10^{6}     \;$ &$20     \;$ &$27     \;$ &$1.2\times 10     \;$ &$1.4\times 10     $ &$1.9\times 10^{-1}$ 	     \\ \hline
$10^{7}     \;$ &$15     \;$ &$22     \;$ &$1.2\times 10^{2} \;$ &$1.4\times 10^{2} $ &$1.9$  	     \\ \hline
$10^{8}     \;$ &$10     \;$ &$17     \;$ &$1.2\times 10^{3} \;$ &$1.4\times 10^{3} $ &$1.9\times 10$  	     \\ \hline
$10^{9}     \;$ &$5.4    \;$ &$12     \;$ &$1.2\times 10^{4} \;$ &$1.4\times 10^{4} $ &$1.9\times 10^{2}$  	     \\ \hline
$10^{10}    \;$ &$0.42   \;$ &$7.3    \;$ &$1.2\times 10^{5} \;$ &$1.4\times 10^{5} $ &$1.9\times 10^{3}$  	     \\ \hline
$10^{11}    \;$ &$-4.6   \;$ &$2.3    \;$ &$1.2\times 10^{6} \;$ &$1.4\times 10^{6} $ &$1.9\times 10^{4}$  	     \\ \hline
\end{tabular}
\end{center}
\end{table}

\section{Discussion and conclusion}
One can detect retrolensing light curves by an Ellis wormhole with a throat parameter $a\geq 10^{7}$km, which is 
within $D_{OL}=1$pc from an observer with current instruments 
if the observer is lucky.
The apparent magnitude $m$ of the peak of a retrolensing light curve by an Ellis wormhole with $a=10^{7}$km at $D_{OL}=1$pc becomes $m=16$ 
if the Ellis wormhole, the observer, and the sun are aligned perfectly in this order.  

It is, however, difficult to distinguish between an Ellis wormhole and a Schwarzschild black hole with the shape and the magnitude of the retrolensing light curves 
made by a pair of light rays with the deflection angle $\alpha \sim \pi$, or with the winding number $N=0$, reflected by a light sphere~\cite{Tsukamoto:2016oca,Tsukamoto:2016zdu}.
In this paper, we have investigated the effect of another pair of light rays with the deflection angle $\alpha \sim 3\pi$, or with the winding number $N=1$, 
reflected by the light sphere on retrolensing to distinguish between an Ellis wormhole and a Schwarzschild black hole.

If a proper length around the light sphere of a lens object is large enough, a retrolensing light curve has a second small peak. 
Otherwise, the second peak cannot be detected because the second peak hides in the first high peak.
The critical proper length around the light sphere depends on 
the ratio between the magnifications of the light rays with $N=0$ and $N=1$; in the other words,
the parameters $\bar{a}$ and $\bar{b}$ of the deflection angles in the strong deflection limit are determined by the details of the line element in a given spacetime.
The proper length $2\pi a$ around the light sphere of an Ellis wormhole becomes critical when the throat parameter $a$ is nearly $10^{11}$km.

An observer can separate a retrolensing light curve into two light curves made by light rays with $N=0$ and $N=1$ 
if the observer detects the second peak of the superposed light curve.
If the observer measures the differences of the apparent magnitudes very precisely,
the observer can say whether the lens object is an Ellis wormhole or a Schwarzschild black hole with the difference of the apparent magnitudes in principle.
In general, the difference of the apparent peak magnitudes can be a strong tool to distinguish given spacetimes with a light sphere.
When an Ellis wormhole with $a=10^{11}$km exists at $D_{OL}=100$pc and the observer and the sun are aligned perfectly, 
the observer measures the apparent magnitude $m_{0}=11$ and $m_{1}=18$ of a pair of light rays with $N=0$ and $N=1$, respectively.
The two pairs of light rays make a superposed light curve with two separable peaks which can be detected with current instruments.
Unfortunately, the difference of the apparent magnitudes of two peaks $m_{0}-m_{1}$ in an Ellis wormhole spacetime 
is almost the same as the difference in the Schwarzschild spacetime 
since the parameter $\bar{a}$ of the deflection angle in the strong deflection limit 
which mainly decides the difference of the apparent peak magnitudes $m_{0}-m_{1}$
in the Ellis wormhole spacetime is the same as the parameter $\bar{a}$ in the Schwarzschild spacetime,

The retrolensing system which we have considered has three unknown parameters: the closest separation $\beta_{m}$,  
the distance between the observer and the lens object $D_{OL}$,
and the throat parameter $a$ if the lens is an Ellis wormhole 
or the ADM mass $M$ if it is a Schwarzschild black hole. 
Since the time delay $\Delta T$ of the light rays with $N=1$ from the light rays with $N=0$ is determined by the proper length around the light sphere,
the observer can determine the throat parameter $a$ of the Ellis wormhole or the ADM mass $M$ of the Schwarzschild black hole 
with a period between a first peak and a second peak of a retrolensing light curve.
From the image separation $\Delta \theta_{0}$ of the double image with $N=0$, 
the observer can determine the distance between the observer and the lens $D_{OL}$.
From the shape of the first peak, the observer can determine the closest separation $\beta_{m}$. 
The observer can determine whether the lens is an Ellis wormhole or a Schwarzschild black hole with the apparent magnitude $m_{0}$ of the first peak of the light curve.
Using the observed apparent magnitude $m_{1}$ of the second peak, 
the observer can confirm the result or determine an extra parameter of the lens object such as an electrical charge.
If the observer measures the image separation $\Delta \theta_{01}$ between the images $N=0$ and $N=1$ on the same side of the lens object,
the observer can confirm the result again or determine another additional parameter of the lens object.  

In this paper, we have concentrated on retrolensing by an Ellis wormhole and a Schwarzschild black hole, 
but the method suggested in this paper could be applied for any retrolens with a large light sphere near the solar system.

\section*{Acknowledgements}
The author is deeply grateful to Ken-ichi Nakao for valuable comments.
He would like to thank Hideki Asada, Yungui Gong, Tomohiro Harada, Takahisa Igata, Masashi Kimura, Takafumi Kokubu, Rio Saitou, Tetsuya Shiromizu, Yusuke Suzuki, 
Yoshimune Tomikawa, Chul-Moon Yoo, and Hirotaka Yoshino for useful conversations. 
The author acknowledges support for this work by the Natural Science Foundation of China under Grant No. 11475065.
\appendix

\end{document}